%% file: tmi.tex
\documentclass[journal,twoside,web]{IEEEtran}
\usepackage{tmi}
\usepackage{cite}
\usepackage{amsmath,amssymb,amsfonts}
\usepackage{siunitx}
\usepackage{mathtools}
\usepackage{algorithm}
\usepackage{algpseudocode}
\usepackage[pdftex]{graphicx}
\usepackage{textcomp}
\usepackage{bm}
\usepackage{hyperref}

\def\min{\mathop{\mathsf{min}}}

\usepackage{diagbox}
\usepackage{multirow}
\usepackage{makecell}
\usepackage{float}
\usepackage{ulem}

\UseRawInputEncoding
\def\BibTeX{{\rm B\kern-.05em{\sc i\kern-.025em b}\kern-.08em
    T\kern-.1667em\lower.7ex\hbox{E}\kern-.125emX}}
\begin{document}
\title{Noise-aware Dynamic Image Denoising and Positron Range Correction for Rubidium-82 Cardiac PET Imaging via Self-supervision}
\author{
Huidong Xie$^{1}$, Liang Guo$^{1}$, Alexandre Velo$^{2}$, Zhao Liu$^{2}$, Qiong Liu$^{1}$, Xueqi Guo$^{1}$, Bo Zhou$^{1}$, Xiongchao Chen$^{1}$, Yu-Jung Tsai$^{2}$, Tianshun Miao$^{2}$, Menghua Xia$^{2}$, Yi-Hwa Liu$^{5}$, Ian S. Armstrong$^{3}$, Ge Wang$^{4}$, Richard E. Carson$^{1,2}$, Albert J. Sinusas$^{1,2,5}$, Chi Liu$^{1,2}$.
\thanks{Corresponding author: Chi Liu.}
\thanks{Emails: \{Huidong.Xie; Chi.Liu\}@yale.edu}
\thanks{$^1$Department of Biomedical Engineering, Yale University, USA.}
\thanks{$^2$Department of Radiology and Biomedical Imaging, Yale University, USA.}
\thanks{$^3$Department of Nuclear Medicine, University of Manchester, UK.}
\thanks{$^4$Department of Biomedical Engineering, Rensselaer Polytechnic Institute, USA.}
\thanks{$^5$Department of Internal Medicine (Cardiology), Yale University, USA.}}
\maketitle

\begin{abstract}
Rubidium-82 ($^{82}\text{Rb}$) is a radioactive isotope widely used for cardiac PET imaging. Despite numerous benefits of $^{82}\text{Rb}$, there are several factors that limits its image quality and quantitative accuracy. First, the short half-life of $^{82}\text{Rb}$ results in noisy dynamic frames. Low signal-to-noise ratio would result in inaccurate and biased image quantification. Noisy dynamic frames also lead to highly noisy parametric images. The noise levels also vary substantially in different dynamic frames due to radiotracer decay and short half-life. Existing denoising methods are not applicable for this task due to the lack of paired training inputs/labels and inability to generalize across varying noise levels. Second, $^{82}\text{Rb}$ emits high-energy positrons. Compared with other tracers such as $^{18}\text{F}$, $^{82}\text{Rb}$ travels a longer distance before annihilation, which negatively affect image spatial resolution. Here, the goal of this study is to propose a self-supervised method for simultaneous (1) noise-aware dynamic image denoising and (2) positron range correction for $^{82}\text{Rb}$ cardiac PET imaging. Tested on a series of PET scans from a cohort of normal volunteers, the proposed method produced images with superior visual quality. To demonstrate the improvement in image quantification, we compared image-derived input functions (IDIFs) with arterial input functions (AIFs) from continuous arterial blood samples. The IDIF derived from the proposed method led to lower AUC differences, decreasing from 11.09\% to 7.58\% on average, compared to the original dynamic frames. The proposed method also improved the quantification of myocardium blood flow (MBF), as validated against $^{15}\text{O-water}$ scans, with mean MBF differences decreased from 0.43 to 0.09, compared to the original dynamic frames. We also conducted a generalizability experiment on 37 patient scans obtained from a different country using a different scanner. The presented method enhanced defect contrast and resulted in lower regional MBF in areas with perfusion defects. Lastly, comparison with other related methods is included to show the effectiveness of the proposed method.

\end{abstract}

\begin{IEEEkeywords}
Enter about five key words or phrases in alphabetical order, separated by commas.
\end{IEEEkeywords}

\input{sec/intro} 
\input{sec/methods}
\input{sec/results}
\input{sec/conclusion}

\bibliographystyle{ieeetr}
\bibliography{tmi.bib}

\end{document}

%% file: sec/intro.tex
\section{Introduction}
\label{sec1}
Positron Emission Tomography (PET) is a functional imaging modality widely used in cardiology studies \cite{gould1991pet, schwaiger2005pet, schindler_cardiac_2010}. Cardiac PET imaging plays a vital role in assessing myocardial perfusion, and ventricular function in patients with known or suspected cardiovascular diseases \cite{ahmed_nuclear_2023}. PET myocardial perfusion imaging with tracer kinetic modeling allows us to quantify regional myocardial blood flow (MBF) and myocardial flow reserve (MFR) of the left ventricle. PET quantitative characteristics provide an objective and more accurate measure of cardiac function than visual inspection alone \cite{schindler_cardiac_2010, boellaard_standards_2009}. Studies have shown that the non-invasive quantification of MBF and MFR offers a predictive measure of cardiovascular diseases \cite{schindler_positron_2005, herzog_long-term_2009,tio_comparison_2009}.

Rubidium-82 ($^{82}\mathrm{Rb}$) is a perfusion PET tracer widely used for cardiac PET imaging in clinical settings \cite{dunet_myocardial_2016}. Compared with myocardial perfusion Single Photon Emission Computes Tomography (SPECT) tracers (e.g., $^{99m}\mathrm{Tc}$-Sestamibi), $^{82}\mathrm{Rb}$ has higher myocardial extraction fraction, allowing a more accurate image quantification \cite{ghotbi_review_2014}. Compared with other perfusion PET tracers (e.g., $^{15}\mathrm{O}$-Water, $^{13}\mathrm{N}$-Ammonia), despite its lower myocardial extraction fraction, $^{82}\mathrm{Rb}$ is generator-produced and does not require an on-site cyclotron \cite{maddahi_cardiac_2014}, making it easily accessible for routine clinical use. $^{82}\mathrm{Rb}$ PET scans also have low effective dose due to its short half-life ($\sim$ 75 seconds). The short half-life also enables fast sequential and repeated scans (e.g., rest and stress scans), improving patient throughput.

Despite numerous advantages of $^{82}\mathrm{Rb}$ for cardiac imaging, there are several physical factors that negatively affect image quality and its quantitative accuracy.

First, dynamic PET imaging measures 4-D spatiotemporal distribution of radioactive tracer in the living body and is essential for tracer kinetic modeling as well as quantification of MBF and MFR \cite{wang_pet_2020}. But the short half-life of $^{82}\mathrm{Rb}$ results in noisy reconstruction of dynamic frames, leading to sub-optimal image quality and quantification results. In addition, compared to tracer kinetic modeling based on a volume of interest (VOI), voxel-wise parametric imaging is more informative and has greater clinical potential \cite{kotasidis_advanced_2014, wang_pet_2020, gallezot_parametric_2020}. Parametric imaging is the process of reconstructing 3-D images of pharmacokinetic parameters from 4-D dynamic SPECT/PET images. However, parametric imaging suffers even more from image noise, especially for fast-decaying tracers like $^{82}\mathrm{Rb}$. Traditional noise-reduction techniques have been utilized to obtain improved parametric images, such as Gaussian smoothing, Bilateral filtering \cite{bian_dynamic_2014}, and Wavelet transforms \cite{alpert_optimization_2006} in the spatial domain. Nonetheless, these methods fail to produce satisfactory results, and better noise reduction techniques for dynamic images are needed \cite{wang_pet_2020}. 

Recently, deep learning has shown great potential for PET image denoising \cite{ouyang_ultra-low-dose_2019, xie_unified_2024, xie_dose-aware_2024, zhou2023fedftn, xu_200x_2017, zhou2020supervised, zhou2023fast, zhou_federated_2023, gong_pet_2019, onishi_anatomical-guided_2021, liu_pet_2022}. However, to the best of our knowledge, current techniques cannot be directly applied to dynamic cardiac PET image denoising. Two problems need to be addressed. First, most of the previously-proposed methods require paired training inputs/labels, which are not feasible to obtain in dynamic $^{82}\mathrm{Rb}$ images due to its short half-life. Lower-noise static frames could be used as pseudo-label or denoised prior for dynamic PET denoising \cite{hashimoto_dynamic_2019}. However, in the case of $^{82}\mathrm{Rb}$ dynamic cardiac PET imaging, the tracer distributions vary substantially between early and later frames, making such technique infeasible for our problem. Unsupervised or self-supervised techniques such as deep-image-prior (DIP) \cite{hashimoto_4d_2021} could be used for dynamic PET denoising. But DIP-based techniques require subject-specific re-training, which is time-consuming and difficult to implement in clinical settings. Other techniques such as noise-to-void (N2V) \cite{krull_noise2void_2019} could also be extended for dynamic PET denoising. However, both DIP-based and N2V methods do not consider the changes in noise-levels and temporal information between different dynamic frames, leading to sub-optimal performance for dynamic PET image denoising, as demonstrated in comparison results included in Section \ref{sec3.4}). Our previous work \cite{xie_unified_2024} proposed to combine multiple sub-networks with varying denoising power to produce optimal denoised results for different input noise levels. But this is a supervised method. In this paper, extended on previous works, we proposed a self-supervised method for for $^{82}\mathrm{Rb}$ dynamic cardiac PET image denoising to consider noise-level and temporal changes between different dynamic frames.

Second, positron range is another physical factor that limits PET image resolution. Positron emission energies are relatively low for the most commonly used radionuclide $^{18}\mathrm{F}$, which has a mean positron range of 0.64 mm in water \cite{conti_physics_2016, garcia_principles_2013}, compared to 1.32 mm for $^{13}\mathrm{N}$, 2.01 mm for $^{15}\mathrm{O}$, and 4.29 mm for $^{82}\mathrm{Rb}$ \cite{garcia_principles_2013}. Higher energy of emitted positrons lead to longer average positron range and thus lower image resolution. Therefore, positron range correction (PRC) is important for $^{82}\mathrm{Rb}$ to enhance image resolution for improved visual assessment and tracer kinetic modeling results. Similar to the dynamic image denoising problem, paired training labels are difficult and time-consuming to obtain for this task. A self-supervised method is also needed.

The positron range distribution can be modeled using Monte Carlo simulations. To overcome the limitations of positron range, the most straightforward approach is Fourier domain division. However, division in the frequency space by a function with low amplitude at high frequencies will enhance high frequencies in the quotient, thus increasing statistical noise \cite{haber_application_1990}.  Previous works also try to incorporate the simulated positron range distributions as an additional point-spread-function (PSF) into the iterative reconstruction updates \cite{bertolli_pet_2016, fu_residual_2010, cal-gonzalez_tissue-dependent_2015, kertesz_implementation_2022}. However, the convergence of these methods is hard to be optimized. Alternatively, modeled positron range distributions can be applied to PET images directly as an image de-convolution using iterative algorithms (e.g., Richardson-Lucy method \cite{richardson_bayesian-based_1972, lucy_iterative_1974}). But because positron range distributions have a blurring effect, iterative de-convolutional methods will inevitably further enhance image noise in dynamic frames. Herraiz \textit{et al.,} \cite{herraiz_deep-learning_2021} proposed a deep learning method for positron range correction by generating paired inputs/labels using simulated emission images from mouse phantoms for supervised network training; this approach is not feasible to translate into clinical settings. Because of these difficulties, positron range correction is not yet adopted for routine clinical use. In addition, most of the previous literature were evaluated only in phantom or small animal studies. Positron range correction on human scans has not been widely explored, especially in the case of parametric imaging and tracer kinetic modeling.

To address the above-mentioned challenges, we propose a self-supervised framework to achieve both (1) noise-aware and temporal-aware dynamic image denoising and (2) positron range correction for $^{82}\mathrm{Rb}$ cardiac PET imaging for improved visual image quality, image quantification, and parametric imaging results. The proposed method was evaluated on a cohort of normal human scans and also clinical patient scans. We conducted a generalizaibilty experiment to show that, without further network fine-tuning, the proposed method could be transferred to patient data of a different population acquired in a different hospital with a different clinical protocol, and scanner, though further validation is needed to show the clinical impact. The proposed method also produced images with potentially improved MBF quantification, as validated against MBF values obtained from $^{15}\text{O-water}$ scans. MBF measurements obtained from $^{15}\text{O-water}$ scans could be considered as a non-invasive reference for MBF quantification as it is almost freely diffusible across capillary and cell membranes \cite{dekemp_toward_2023, manabe_15o-labeled_2019}, with single-pass extraction fraction close to 1 \cite{bergmann_quantification_1984}. But $^{15}\text{O-water}$ requires an on-site cyclotron, has not yet been adopted for clinical use, and is not ideal for visual assessment \cite{dekemp_toward_2023, manabe_15o-labeled_2019}. Lastly, the proposed method produced images with improved image quantification, as compared against radio-activities quantify with continuously-measured arterial blood samples.

%% file: sec/methods.tex
\section{Methodology}
\subsection{Data Acquisition and Image Reconstructions}
The proposed method was evaluated on dataset acquired on a Siemens Biograph mCT PET/CT system at Yale PET Center during rest and under pharmacological stress (induced with 0.4 mg of regadenoson) with $^{82}\mathrm{Rb}$ and $^{15}\mathrm{O}$-water for each subject \cite{germino_quantification_2016}. Cardiac PET studies from a total of 9 normal volunteers (five male) with no known cardiac abnormalities were included. The average age was $28.4\pm6.2$ years, and average BMI was $24.7\pm3.9$ $\mathrm{kg/m^2}$. There was roughly a 1-hour separation between stress and rest scans, with confirmation that the heart rate and blood pressure had returned to baseline. For attenuation correction, low-dose CT scans were performed before each rest scan and after each stress scan. For all subjects, $\text{mean}\pm \text{SD}$ of injection dose were $663\pm82 \text{MBq}$ for $^{82}\mathrm{Rb}$, and $690\pm316 \text{MBq}$ for $^{15}\mathrm{O}$-water. Contrast-enhanced CT scans were performed for some of the normal volunteers. Scan duration was 6 minutes from the time of injection for each subject. List-mode data were reconstructed into 38 dynamic frames ($20\times3s,6\times10s,12\times20s$) with TOF (Time of flight) information, PSF modeling, and prompt-gamma corrections for $^{82}\mathrm{Rb}$ studies. Images were reconstructed using OSEM (ordered subset expectation maximization) \cite{hudson_accelerated_1994} with 2 iterations of 21 subsets. A 3 mm-FWHM Gaussian post-filtering was applied. The reconstructed matrix size was $400\times400\times109$ with $2.036\  \text{mm}\times2.036\ \text{mm}\times2.0\ \text{mm}$ voxel size. Static frame images were separately reconstructed using list-mode data from 120s to 360s.

To non-invasively quantify MBF and MFR, input functions derived from the dynamic PET images (i.e., image-derived input function, IDIF) were used. But PET images may subject to quantification bias, resulting in inaccurate measurements of MBF and MFR. To show that the proposed denosing and positron range correction method improve image quantification, arterial blood was collected and radioactivity was quantified as a gold-standard for comparison. Seven of the nine subjects chose to perform arterial blood sampling during the scans. Arterial blood was drawn from the radial arterial for 7 minutes per scan at 4 mL per minute. Radioactivity was measured with a cross-calibrated radioactivity monitor (PBS-101, Veenstra Instruments). IDIFs can then be compared with AIFs as an additional validation for improved image quantification. Further data acquisition details are available in our previous publication \cite{germino_quantification_2016}.

Since the positron range effect should be independent of the scanner, we also evaluated the generalizability of the proposed positron range correction method using 37 patient scans obtained on a different scanner (Siemens Biograph Vision PET/CT) at the University of Manchester Hospital. Scan duration was 5 minutes for each subject. 35 dynamic frames were reconstructed ($20\times3s,6\times10s,9\times20s$) with TOF, PSF, and prompt-gamma corrections using OSEM with 3 iterations of 5 subsets. Reconstructed matrix size was $440\times440\times109$ with $1.65\ \text{mm}\times1.65\ \text{mm}\times1.65\ \text{mm}$ voxel size. A 3 mm-FWHM Gaussian post-filtering was applied. Further data acquisition details are available in \cite{armstrong_preliminary_2022}.

All the images were reconstructed using vendor's software from Siemens Healthineers.

\subsection{Proposed Deep-learning Framework}

\begin{figure*}[!t]
\centerline{\includegraphics[width=\textwidth]{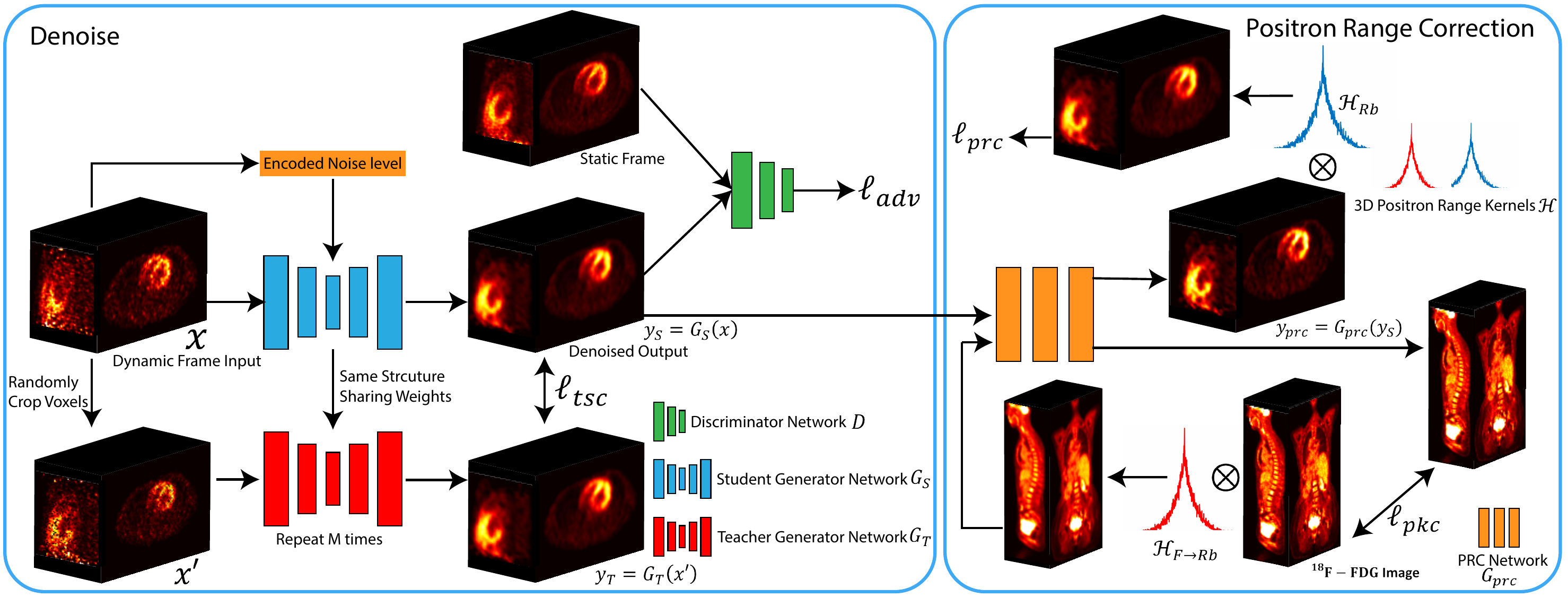}}
\caption{The proposed deep-learning framework for 3-D self-supervised noise-ware dynamic image denoising and positron range correction (PRC). It can be divided into 2 components. One for dynamic image denoising and the other for PRC. Dynamic frames first go through the denoising component and then the PRC component to achieve both dynamic image denoising and PRC.}
\label{fig1}
\end{figure*}

\begin{figure}[!t]
\centerline{\includegraphics[width=\linewidth]{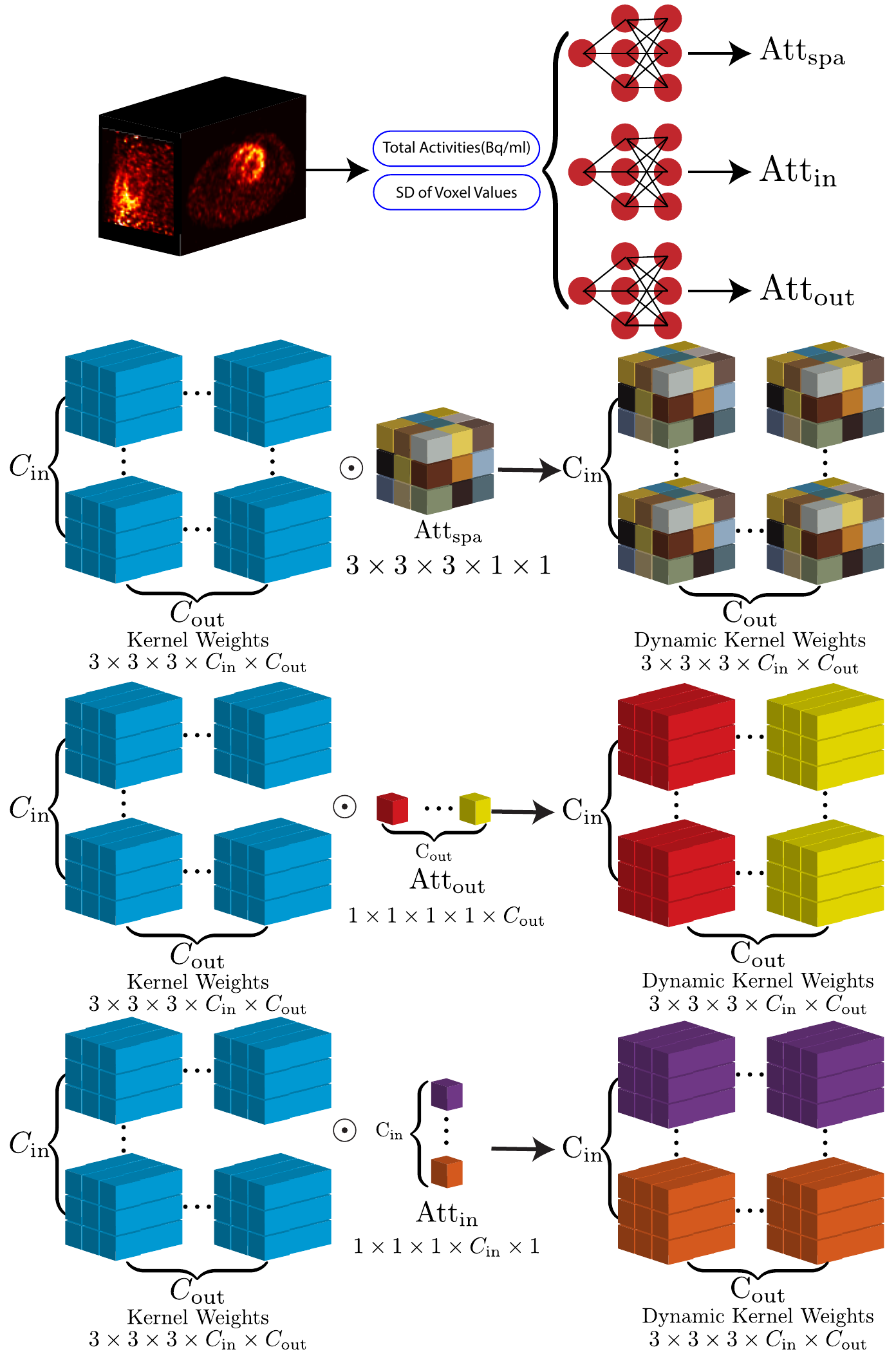}}
\caption{Graphical illustration of the proposed dynamic convolutional strategy with kernel size $3\times3\times3$ as an example. Three attention weights $\text{Att}_{\text{spa}}$, $\text{Att}_{\text{in}}$, and $\text{Att}_{\text{out}}$ are obtained using the encoded noise information. $C_{in}$ and $C_{out}$ indicate input channel dimension and output channel dimension respectively. Blue cubes represent the values of the convolutional kernel before applying the dynamic attention weights. Non-blue colors represent how the dynamic attention weights are applied. Three dynamic kernel weights are then averaged before performing the convolutional operations.}
\label{fig2}
\end{figure}

The overall proposed framework is depicted in Fig. \ref{fig1}. The proposed neural network consists of 2 components, one for dynamic image denoising and the other for positron range correction. The 3-D dynamic images are first fed into the denoising component to produce lower-noise images and then fed into the PRC component to achieve positron range correction.

\subsubsection{Self-supervised Noise-aware Dynamic Image Denoising}
Given the noisy dynamic frames $x\in \mathbb{R}^{W\times W\times D}$ as input, the goal of the denoising component is to denoise dynamic frames so that the noise level is similar to static frame reconstructions. $W$ and $D$ represent the width and depth of the reconstructed matrix size. To enforce the similarity of the noise levels, the Wasserstein Generative Adversarial Network (WGAN) with gradient penalty \cite{NIPS2017_892c3b1c} was implemented in the denoising component. The WGAN architecture contains 2 separate networks, one generator network $G$ aims to denoise dynamic frames, and the other discriminator network $D$ aims to distinguish the fidelity of the input (either generated from $G$ or from static frame list-model data). Throughout the training process, the generator network $G$ will tend to generate denoised images that are close to static frame in terms of overall noise level. As presented in Fig. \ref{fig1}, the adversarial loss $\ell_{adv}$ was included for network optimization.

To achieve self-supervised dynamic image denoising, the proposed denoising method builds from the Noise2Void (N2V) \cite{krull_noise2void_2019} idea. N2V has demonstrated successful implementations for medical image denoising \cite{song_noise2void_2021}. Inspired by the N2V idea, roughly 50\% of voxels in the images were randomly removed to generate $x'$ in Fig. \ref{fig1}. Note that the majority of the voxels are zeros in the entire image volume. Partially-cropped images $x'$ are then fed into the neural network using the original noisy input values as training targets. The N2V approach involves training a network using identical noisy input and target. In this circumstance, the network will tend to generate an output that is the same as the input. To prevent the network from learning the identity, N2V uses a blind-spot design that masks out certain voxels in the image volume, encouraging the network to seek information from neighboring voxels, achieving image denoising, as the image signals are spatially correlated.

To prevent the network from generating unrealistic features in the cropped regions, the mean teacher model \cite{tarvainen_mean_2017, xia_multilevel_2022} was adapted to generate voxel-wise pseudo label as an additional constrain to the denoised output. As presented in Fig. \ref{fig1}, the denoising network contains 2 generator networks, namely the student generator network $G_S$, and the teacher generator network $G_T$. Both $G_S$ and $G_T$ share the same network structure. The input to the network $G_T$ is the partially-cropped image $x'$ (i.e., dynamic frames $x$ with cropped voxels). The input to the network $G_S$ is the original dynamic frames $x$. Within each training step $t$, the teacher network ($G_T$) parameters $\theta_T$ is the exponential moving average of the student network ($G_S$) parameters $\theta_S$:

\begin{equation}
\theta_T(t) = \alpha \theta_T(t-1) + (1-\alpha) \theta_S(t)
\label{eq1}
\end{equation}

where $\alpha=0.99$ is a hyperparameter that controls the parameter update rate. 

To generate a pseudo label for network training, $M$ different partially-cropped images ($x_m', m=1,...,M$) were generated and fed into $G_T$. The final prediction of the teacher network $G_T$ is defined as the mean of $M$ different stochastic forward passes of $G_T$:

\begin{equation}
\hat{y}_{T}=\frac{1}{M}\sum_{m=1}^MG_T(x_m')
\label{eq2}
\end{equation}

The uncertainty $u$ of all the $M$ predictions is defined as:

\begin{equation}
u=\frac{1}{M}\sum_{m=1}^M(\hat{y}_{T}-G_T(x_m'))
\label{eq3}
\end{equation}

Here, $\hat{y}_{T}$ is considered as a pseudo for the student network $G_S$. The prediction reliability of each voxel $i$ is quantified by the uncertainty term $u(i)$. In Fig. \ref{fig1}, the teacher-student consistency loss function $\ell_{tsc}$ is designed so that voxels with higher uncertainties have lower weights in the loss function and vice versa. To achieve this, $\ell_{tsc}$ is formulated as:

\begin{equation}
\ell_{tsc} = \frac{\sum_i[1-u(i)]|\hat{y}_{T}(i)-y_S(i)|}{\sum_i[1-u(i)]}
\label{eq4}
\end{equation}

where $y_S=G_S(x)$ represents the output from the student network $G_S$.

To consider the noise-level differences across different dynamic frames and achieve noise-aware denoising, the noise-level information is encoded into the neural network using the idea of dynamic convolution \cite{xie_segmentation-free_2023, li2022omnidimensional}. Convolutional-based networks attempt to learn static convolutional kernels during the training process, and the learned kernels are fixed in the testing phase. In the case of dynamic convolution, a set of attention weights are obtained from the input features and applied to different dimensions of the convolutional kernel, thus improving the generalizability of the network to different input noise levels. Our previous work presented a successful implementation of dynamic convolution for cardiac SPECT partial volume correction \cite{xie_segmentation-free_2023}. In this work, we extended the idea of dynamic convolution to achieve noise-aware denoising.

A graphical illustration of the proposed dynamic convolution strategy is presented in Fig. \ref{fig2}. The 3-D convolutional operation can be formulated as:

\begin{equation}
\mathcal{F}_{out} = \mathcal{W}\otimes \mathcal{F}_{in} + \mathcal{B}\label{eq5}
\end{equation}

where $\mathcal{F}_{in}\in\mathbb{R}^{d\times w\times h\times C_{in}}$, and $\mathcal{F}_{out}\in\mathbb{R}^{d\times w\times h\times C_{out}}$ represent input and output feature maps, respectively. $d$, $w$, and $h$ denote the spatial dimension of the input/output feature maps, which may be different based on the parameters of the convolutional layer. $C_{in}$ and $C_{out}$ are the input and output channel dimensions. $\mathcal{W}\in\mathbb{R}^{k\times k\times k\times C_{in} \times C_{out}}$ denotes the convolutional kernel weights, and $\mathcal{B}\in\mathbb{R}^{C_{out}}$ is the bias term. $k$ is the spatial dimension of the convolutional kernel. $\otimes$ represents the convolutional operator.

In the proposed dynamic convolutional strategy, the kernel weights $\mathcal{W}$ become adaptive based on the encoded noise information. We used total activities in Bq/ml and the standard deviation of the non-zero voxel values as indicators of image noise level. $\sin$ and $\cos$ functions were used for encoding. Specifically, $\text{encoding}=sin(\text{total activities}) + cos(\text{SD of voxel values})$. The encoded values are then fed into three sets of 2 dense layers to generate 3 attention weights, $\text{Att}_{\text{spa}}\in\mathbb{R}^{k\times k\times k}$, $\text{Att}_{\text{in}}\in\mathbb{R}^{C_{in}}$, and $\text{Att}_{\text{out}}\in\mathbb{R}^{C_{out}}$. Rectified linear unit (ReLU) and sigmoid are used as the activation functions after the first and the second dense layers, respectively. With the proposed dynamic convolutional strategy, equation (\ref{eq5}) becomes:

\begin{equation}
\mathcal{F}_{out} = [\mathcal{W}\odot \frac{1}{3}(\text{Att}_{\text{spa}}+\text{Att}_{\text{in}}+\text{Att}_{\text{out}})] \otimes \mathcal{F}_{in} + \mathcal{B}\label{eq6}
\end{equation}

To this end, we described the proposed framework to achieve self-supervised noise-aware dynamic image denoising. The composite objective function to optimize the denoising network is formulated as:

\begin{equation}
\underset{{\theta}_S}{\min}\ L_{\text{denoise}} =\bigg\{\ell_{tsc} + \underbrace{\lambda_a\,\mathbb{E}_x\left[D(G_S(x))\right]}_{\text{adversarial loss }\ell_{adv}} + \ell_{\mathrm{MAE}}(x,y_S) \bigg\}
 \label{eq7}
\end{equation}

where $\lambda_a$ is hyper-parameter used to balance different loss functions. $\mathbb{E}_a[b]$ denotes the expectation of $b$ as a function of $a$. The mean-absolute-error $\ell_{\mathrm{MAE}}$ between the input $x$ and the output $y_S=G_S(x)$ was also included for network optimization to prevent the network from generating unrealistic structures.

\subsubsection{Self-supervised Positron Range Correction}
As mentioned previously, positron range distributions can be simulated using the Monte Carlo method. To achieve positron range correction, the network can be designed to learn the reverse of the simulated positron range kernel. In the context of this paper, we assumed the positron range kernel is spatially uniform when training the neural network.

As presented in Fig. \ref{fig1}, to achieve positron range correction, the denoised output $y_S=G_S(x)$ is fed into the positron range correction network $G_{prc}$ to obtain the positron range correction results $y_{prc}=G_{prc}(y_S)$. To learn the inverse of the $^{82}\text{Rb}$ positron range kernel, the network parameters were optimized using the following objective $\ell_{prc}$:

\begin{equation}
\ell_{prc} = \ell_{\mathrm{MAE}}(y_{prc}\otimes\mathcal{H}_{Rb}, y_S)
 \label{eq8}
\end{equation}

where $\mathcal{H}_{Rb}$ represents the simulated positron range kernel of $^{82}\mathrm{Rb}$ using Monte Carlo method. Specifically, because the network $G_{prc}$ is designed to approximate the inverse of $\mathcal{H}_{Rb}$, in the objective function $\ell_{prc}$, the network output $y_{prc}$ is convoluted with $\mathcal{H}_{Rb}$, and the convoluted image is expected to be the same as the network input $y_S$. The MAE between them was used for network optimization.

However, because the positron range kernel $\mathcal{H}_{Rb}$ has a blurring effect, if the network $G_{prc}$ perfectly models the inverse of it, the output $y_{prc}$ is expected to be noisy, which is not desirable. To address this issue, we proposed to use pseudo labels generated using $^{18}\text{F-FDG}$ images. Specifically, pseudo labels were created by simulating $^{82}\mathrm{Rb}$ positron range effects on $^{18}\text{F-FDG}$ images. This was achieved by convoluting $^{18}\text{F-FDG}$ with the kernel $\mathcal{H}_{F\rightarrow rb}$, which models the additional blurring between $^{18}\text{F}$ and $^{82}\mathrm{Rb}$.

\begin{equation}
\mathcal{H}_{Rb} = \mathcal{H}_{F}\otimes \mathcal{H}_{F\rightarrow rb}
 \label{eq9}
\end{equation}

where $\mathcal{H}_{F}$ represents the simulated positron range kernel of $^{18}\text{F}$ using the Monte Carlo method. $\mathcal{H}_{F\rightarrow rb}$ represents the kernel converting $^{82}\text{Rb}$ to $^{18}\text{F}$. Note that $\mathcal{H}_{F\rightarrow rb}$ cannot be directly simulated using the Monte Carlo method. In this work, $\mathcal{H}_{F\rightarrow rb}$ was approximated using gradient descent with mean-absolute-error as the optimization metric between $\mathcal{H}_{F}$ and $\mathcal{H}_{Rb}$. The $^{82}\mathrm{Rb}$ blurred $^{18}\text{F-FDG}$ images were used as the input to the network $G_{prc}$, and the MAE between the network output and the original $^{18}\text{F-FDG}$ images were used for network training. This loss function is depicted as the positron kernel consistency loss ($\ell_{pkc}$) in Fig. \ref{fig1}. Lastly, the MAE between $y_S$ and $y_{prc}$ was also included as an additional constraint to prevent the images from becoming too noisy ($\ell_{idt}$). The composite objective function of the network $G_{prc}$ is formulated as:

\begin{equation}
\underset{{\theta}_{prc}}{\min}\ L_{\text{prc}} = \bigg\{\ell_{prc} + \lambda_b \ell_{idt} + \ell_{pkc}\bigg\}
 \label{eq10}
\end{equation}

where $\theta_{prc}$ represents the trainable parameters of the network $G_{prc}$, $\lambda_b$ is a hyper-parameter used to prevent the identity difference from overwhelming other loss terms.

\subsubsection{Network Structure}
In the denoising component, both networks $G_S$ and $G_T$ share the same structure. They follow a U-net-like structure \cite{ronneberger_u-net_2015}. Both networks consist of four 3-D down-sampling and four 3-D up-sampling convolutional layers. A 3-layer dense-net structure \cite{huang_densely_2017} is added after each down-/up-sampling layer, followed by a squeeze-excite attention block \cite{hu_squeeze-and-excitation_2018}. Note that the proposed dynamic convolutional strategy was implemented in the dense-net blocks. Another 3-D convolutional layer is added at the end of the network to produce one-channel output. All the 3-D convolutional layers used for down-/up-sampling have a kernel size of $3\times 3\times 3$ with a stride of 1 without zero-padding. The 3-D convolutional layers in the dense-net block have a kernel size of $5\times3\times3$ with a stride of 1 and zero-padding. ReLU activation functions are implemented after each layer except the last layer. All the convolutional layers have 32 filters, except the last layer only has 1 filter.

The discriminator network $D$ in the denoising component has six 3-D convolutional layers with 64, 64, 128, 128, 256, and 256 filters and two fully-connected layers with the number of neurons 1024 and 1. The leaky ReLU activation function is added after each layer with a slope of 0.2 in the negative component. Convolution operations are performed with $3\times3\times3$ kernels and zero-padding. Stride equals 1 for odd-numbered layers and 2 for even-numbered layers. 

The positron range correction network $G_{prc}$ consists of five 3-D convolutional layers. All of them have a kernel size of $3\times 3\times 3$ with a stride of 1 and zero-padding. ReLU activation functions are implemented after each layer except the last layer. All the convolutional layers have 32 filters, except the last layer only has 1 filter.

\subsection{Network Optimization and Training}
The network was trained in 2 separate steps. In the first step, the denoising and the positron range correction components were trained separately. The denoising component was trained using dynamic frames and the positron range correction component was trained using static frames. In the second step, the entire framework was fine-tuned in an end-to-end fashion using dynamic frames as input. The network was trained using only the 9 normal volunteers (18 scans, both rest and stress) acquired on a Siemens mCT scanner. To obtain testing results for all the mCT studies, the proposed framework was re-trained 9 separate times. Within each training iteration, one subject was used for testing, one subject was used for validation, and the remaining seven subjects were used for network training. Patch-based training strategy was implemented. In the denoising component, a patch size of $128\times128\times20$ was used. Patches with majority zeros were excluded. In the positron range correction component, a patch size of $360\times360\times20$ was used. Experimental results showed that the denoising component required more training data to converge, so we implemented a smaller patch size to generate more training data. Since ground-truth training labels were not available, $\lambda_a=0.05$ and $\lambda_b=0.5$ were experimentally fine-tuned. The trained network was then directly applied to 37 patient studies acquired on a Siemens Vision PET/CT system.

\subsection{Monte Carlo Simulation Details}
The simulations were performed using the MCNP (Monte Carlo N-Particle) package \cite{forster_mcnp_1985}. 300,000 positrons were simulated in uniform tissues of lung (mass density 0.3 $g/cm^3$), soft tissue (1 $g/cm^3$), skeletal muscle (1.04 $g/cm^3$), and striated muscle (1.04 $g/cm^3$). Material compositions were obtained from the NIST (National Institute of Standards and Technology) database. Human tissues close to the cardiac regions are mainly combinations of these four tissues. Eight simulations were performed for both $^{18}\text{F}$ and $^{82}\text{Rb}$. Average positron range values for different simulations are summarized in Table \ref{table1}. The mean positron range and the distributions are reasonably close in soft tissue, skeletal muscle and striated muscle. The distributions are much wider in the lung due to lower tissue density. In this work, since we focused on cardiac imaging, simulations performed in a uniform tissue of striated muscle was used. $\mathcal{H}_{Rb}$ and $\mathcal{H}_{F}$ were created by interpolating the annihilation end-points based on the image voxel size.

\begin{table}[!h]
\centering
\caption{Simulated mean positron range (mm) for $^{18}\text{F}$ and $^{82}\text{Rb}$ in four different tissues.}
\resizebox{\linewidth}{!}{
\begin{tabular}{c|c|c|c|c}
\hline
   \backslashbox{Isotopes}{Tissues}  & Lung &  Soft Tissue & Skeletal Muscle & Striated Muscle     \\

\hline
$^{18}\text{F}$ & 1.9840 &  0.5967 & 0.5725 & 0.5720  \\
 \hline
$^{82}\text{Rb}$  & 15.3278 & 4.6774 & 4.4876 & 4.4852 \\

\hline

\end{tabular}
}
\label{table1}
\end{table}

\subsection{Tracer Kinetic Modeling and Parametric Imaging}

The three-parameter one-tissue compartment model was used to describe the tracer kinetics in the myocardium. The tissue tracer concentration for a specific voxel or region at time $t$ can be expressed as: 

\begin{equation}
C_\mathrm{T}(t) = V_\mathrm{b} C_\mathrm{b}(t) + (1-V_\mathrm{b})(K_1 e^{-k_2 t}\otimes C_\mathrm{b}(t))
\label{eq11}
\end{equation}

where $K_1$ and $k_2$ are the influx and efflux rates, respectively. $C_\mathrm{b}(t)$ is the image-derived input function from the left ventricle blood pool and  $ C_\mathrm{T}(t)$ represents the time-activity curve of the left ventricular myocardium. $V_b$ stands for fractional blood volume. Regional $K_1$, $k_2$, and $V_b$ values were calculated by averaging all the voxels in the volume of interest (VOI), which were obtained by manual segmentation of the 3-D image volumes. Equation (\ref{eq11}) was fit to each voxel using the basis function method \cite{lodge_parametric_2000} to generate voxel-wise parametric images. The generalized Renkin-Crone model was used to quantify MBF for $^{82}\text{Rb}$ studies \cite{renkin_transport_1959,crone_permeability_1963}.

\begin{equation}
K_1 = \text{MBF}(1-ae^{-b/\text{MBF}})
\label{eq12}
\end{equation}

The parameters $a=0.74$, and $b=0.51$ fitted in our previous work were used \cite{germino_quantification_2016}. The parameters were determined using paired dynamic $^{82}\text{Rb}$ and the $^{15}\text{O-water}$ scans. For $^{15}\text{O-water}$ scans, MBF was estimated from the mean myocardial $^{15}\text{O-water}$ $k_2$ values, corrected with a partition coefficient of $p=0.91 \text{mL/g}$ ($\text{MBF}=k_2p$) \cite{Iida_measurement}. MFR is defined as the ratio between the stress and rest MBF measurements. MFR represents the relative reserve of the coronary circulation, and there is no optimal value for it. Typically, $\text{MFR}>2.3$ indicates a favorable prognosis and $\text{MFR}<1.5$ suggests significantly diminished flow reserve \cite{ziadi_myocardial_2017}.

In this paper, the MBF values quantified using the $^{82}\text{Rb}$ scans with the proposed positron range correction were validated against the MBF obtained from the $^{15}\text{O-water}$ scans with a much smaller positron range. $^{15}\text{O-water}$ offers precise MBF quantification as its has 100\% extraction fraction even at high flow rate. 

IDIFs were estimated using VOI manually determined in the left ventricular blood pool for rest and stress scans for each subject using the $^{82}\mathrm{Rb}$ static frame reconstructions. Cylindrical VOIs were placed along the center of the basal to mid-ventricular cavity. Myocardium VOIs with approximately 2-4 voxels in width ($\sim\text{4-8}$ mm) were placed along the center line of the left ventricle. With a sufficiently small VOI, there is nearly complete recovery of the arterial input curve and minimal myocardial spillover \cite{lortie_quantification_2007}.

%% file: sec/results.tex
\section{Results}

\begin{figure*}[!t]
\centerline{\includegraphics[width=\textwidth]{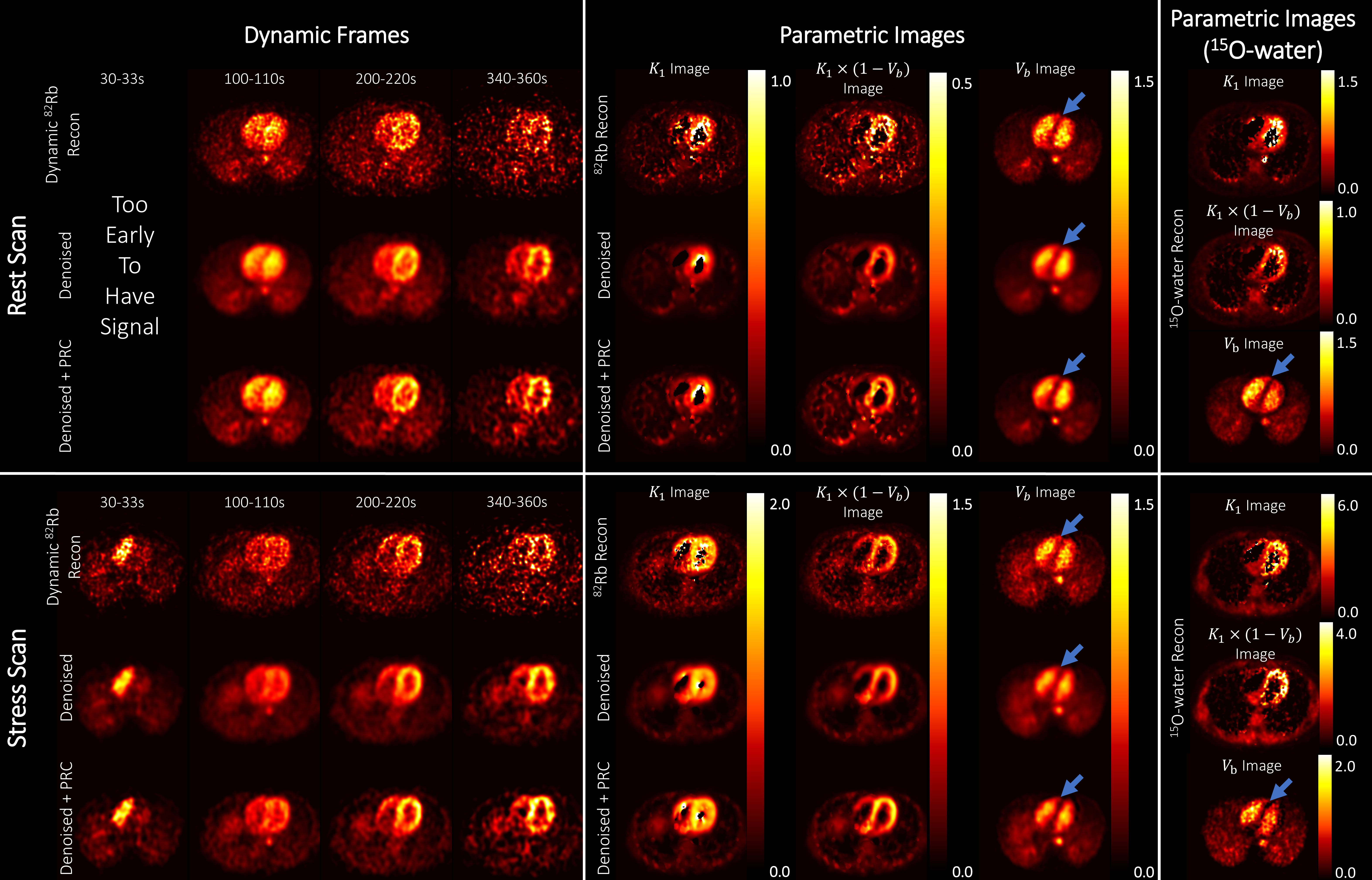}}
\caption{Normal volunteer study obtained using the Siemens mCT PET/CT system. Both rest and stress studies are presented. Dynamic frames from both rest and stress studies are presented. The proposed method can generalize to images with different noise levels. In the rest scan, dynamic interval 30s-33s is too early to have a detectable signal. $K_1$ images are multiplied by $(1-V_b)$ to remove the artifacts at the boundary of myocardium and blood pool for better visualization (denoted as $K_1\times(1-V_b)$ image).}
\label{fig4}
\end{figure*}

\subsection{Visual Observation}
One normal volunteer subject obtained on the Siemens mCT scanner is presented in Fig. \ref{fig4}. The denoising network ($G_S$) produced lower-noise images, and the positron range correction network ($G_{prc}$) produced sharper images with clearer myocardium contour in later frames and blood pool in early frames. The proposed $G_S$ can effectively generalize to dynamic frames with different noise levels and different tracer distributions. The proposed method was able to recover reasonable reconstructions even for the last dynamic frame (340s-360s), in which original list-model data were not able to produce images with clear cardiac contour.

\begin{figure}[!ht]
\centerline{\includegraphics[width=\linewidth]{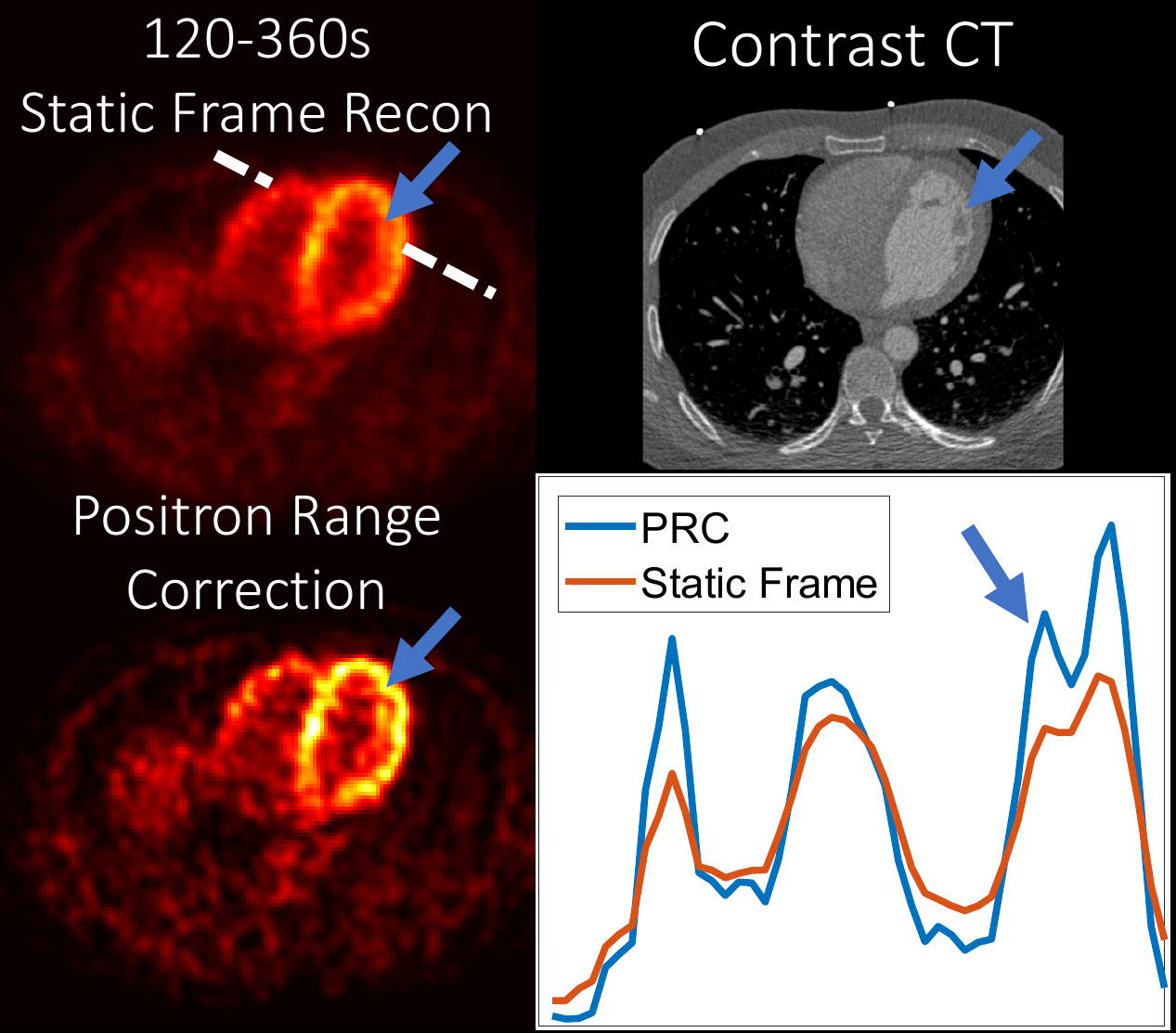}}
\caption{Normal volunteer study obtained using the Siemens mCT PET/CT system. The stress static frame results are included in this figure. Profile plots were generated along the dashed white line. Blue arrows in the profile plots and the images point to the papillary muscle as validated by the contrast CT scan.}
\label{fig5}
\end{figure}

Results from static frames are presented in Fig. \ref{fig5}. Since the goal of the $G_S$ is to denoise dynamic frames so that the noise level aligns with the static frames, static frames do not require denoising. As presented in Fig. \ref{fig5}, in addition to better image resolution, the proposed positron range correction $G_{prc}$ produced images with more subtle features revealed. For example, the papillary muscle pointed by the blue arrows in Fig. \ref{fig5} is better visualized in the positron range correction results, as confirmed by the contrast-enhanced CT scan and the profile plots. These small structures are usually challenging to identify due to limited spatial resolution \cite{nakao_papillary_2022}, especially in $^{82}\text{Rb}$ cardiac PET images. But the proposed $G_{prc}$ produced images with higher resolution and better visualization of these small cardiac structures, confirming the improved image resolution. 

Since there is no ground-truth image for comparisons, we calculated the myocardium-to-blood pool ratios for the static frame results to show the improvement in image contrast. The proposed $G_{prc}$ consistently produced images with higher myocardium-to-blood pool ratios. For stress scans, the numbers are $2.79\pm0.52$ and $3.79\pm0.86$ for static frame inputs and the positron range correction outputs, respectively, representing a $35.24\pm7.16\%$ increase. For rest scans, these numbers are $1.75\pm0.32$ and $2.13\pm0.49$, respectively, representing a $20.89\pm7.44\%$ increase.

\subsection{Tracer Kinetic Modeling and Parametric Imaging}

\begin{figure}[!t]
\centerline{\includegraphics[width=\linewidth]{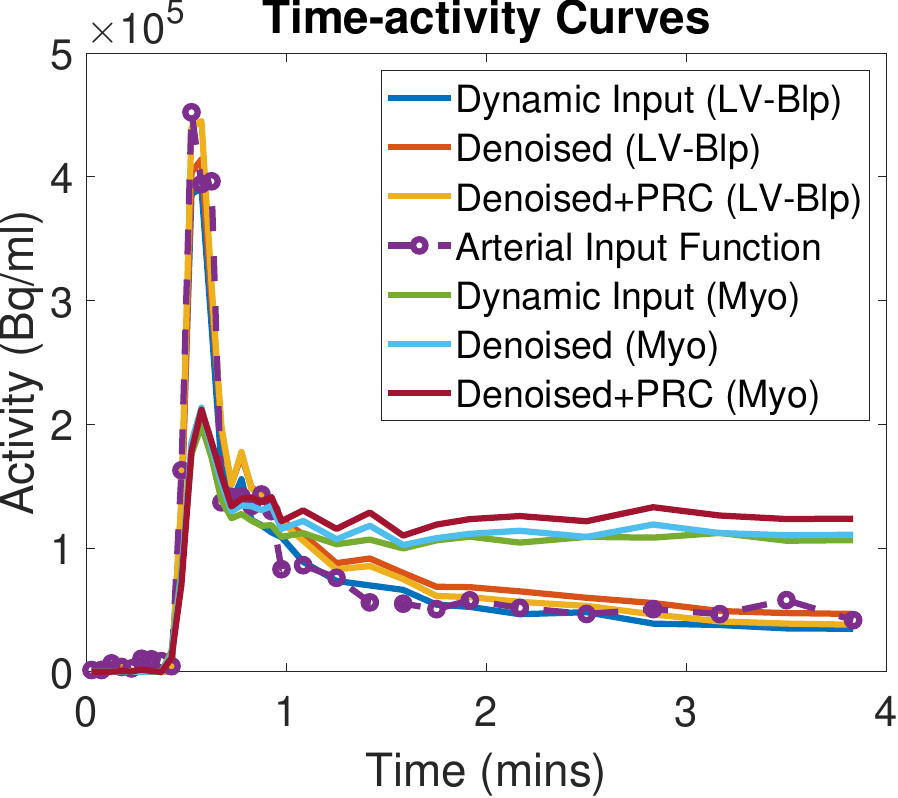}}
\caption{Arterial input function (AIF) and image-derived time-activity curves for different dynamic series from a sample subject acquired on the Siemens mCT PET/CT system. Proposed "denoised+PRC" method produced images with a better match to the AIF. LV-Blp: left ventricular blood-pool; Myo: Myocardium.}
\label{fig7}
\end{figure}

Using the VOIs manually placed in the myocardium and the blood-pool, the resulting time-activities curves were compared to the measured AIF with regard to peak concentration, tail concentration, and area under the curve (AUC). For comparison, AIFs were resampled to the image times by averaging values within each frame. Peak concentrations were computed as the maximal activity of each TAC. Tail concentrations were computed by averaging the concentration between 2.16 min to 4 min post-injection. TAC curves from one of the normal volunteers are presented in Fig. \ref{fig7}. The proposed method produced images with similar peak to the AIF. $G_{prc}$ produced images with higher myocardium activities as the myocardium becomes sharper in the images after positron range correction. The absoluate percentage differences between AIF and image-derived input function (IDIF) are included in Table \ref{table2}. Proposed neural network produced images with TACs better matched with AIFs with a overall lower percentage difference.

Corresponding reconstructed $K_1$ and $V_b$ parametric images are also included in Fig. \ref{fig4}. Due to the high image noise, $K_1$ images derived from the original dynamic frames are very noisy. The proposed $G_S$ produced lower-noise $K_1$ images. The $G_{prc}$ produced sharper $K_1$ images with better myocardium contour. As presented in Table \ref{table2}, due to the smoothing introduced in the denoising network, denoised results have lower average $K_1$ values than the original dynamic frames. In addition, due to lower myocardium influx rate in the rest scan, the rest $K_1$ image is even nosier than the stress $K_1$ image. The proposed method was still able to produce lower-noise $K_1$ image with better myocardium boundaries.

The proposed network also produced lower-noise $V_b$ images. As indicated by the lower mean $V_b$ values, the $V_b$ images produced by the proposed positron range correction method present better separation between the left and right ventricular blood pools. $K_1$ and $V_b$ images from the corresponding $^{15}\text{O-water}$ scans are also included in Fig. \ref{fig4}. Due to shorter positron range of $^{15}\text{O}$, the $V_b$ images derived from $^{15}\text{O-water}$ data also present more clear septal wall between left and right ventricular blood pools compared with the original $^{82}\text{Rb}$ dynamic images. But $^{15}\text{O-water}$ images are still noisy due to the short half-life ($\sim122.3s$).

\begin{figure*}[!t]
\centerline{\includegraphics[width=\linewidth]{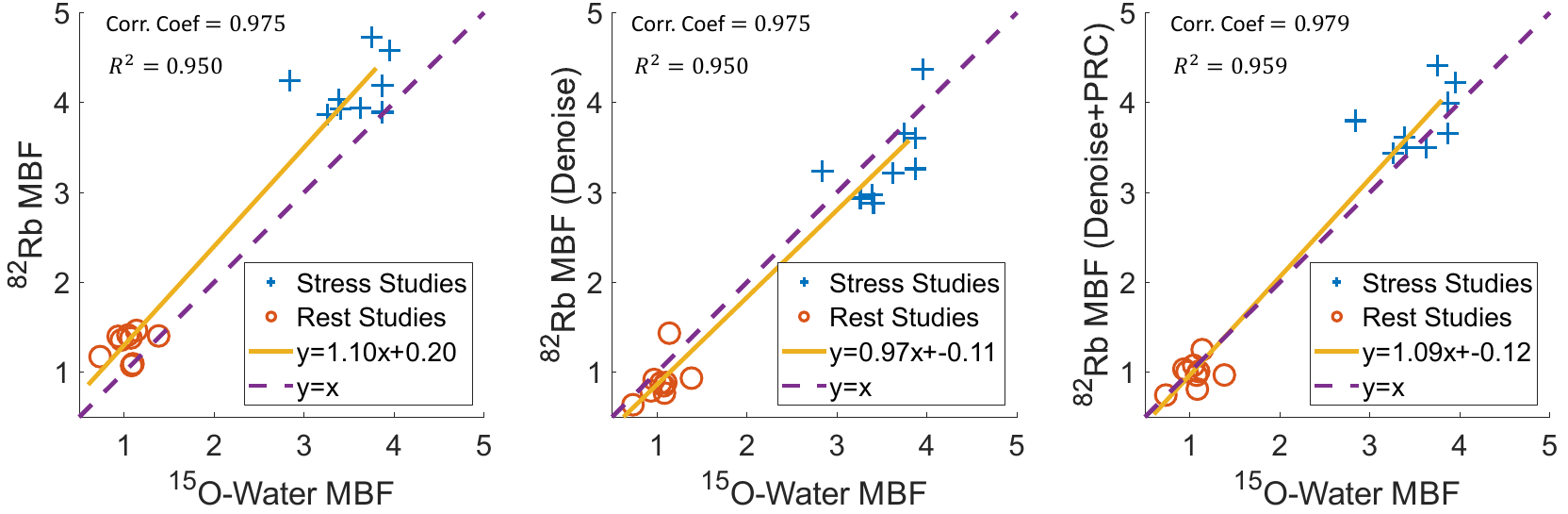}}
\caption{Linear fitting plots for comparing MBF (ml/min/g) calculated from $^{15}\text{O-water}$, and $^{82}\text{Rb}$ images reconstructed using different methods. Proposed method produced images with MBF measurements better align with the $^{15}\text{O-water}$ scans, which served as the reference MBF values in this paper. Coefficient of determination ($R^2$) and the correlation coefficient (Corr. Coef) are included in the plots.}
\label{fig6}
\end{figure*}

\begin{figure*}[!t]
\centerline{\includegraphics[width=\linewidth]{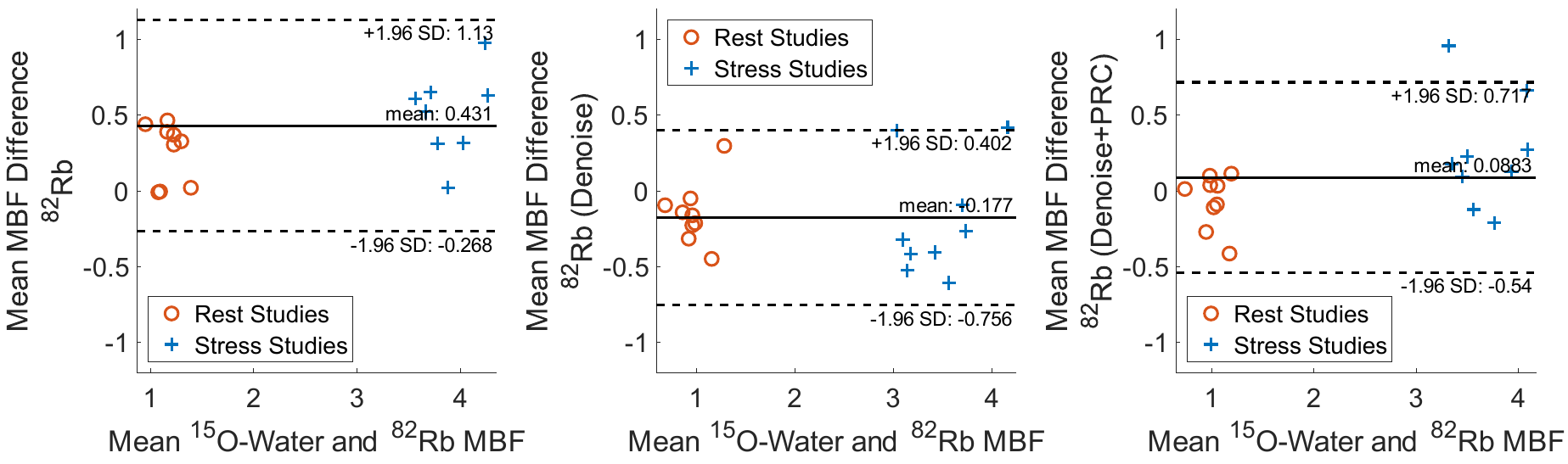}}
\caption{Bland–Altman plots for comparing MBF (ml/min/g) calculated from $^{15}\text{O-water}$, and $^{82}\text{Rb}$ images reconstructed using different methods. Proposed method produced images with MBF measurements better align with the $^{15}\text{O-water}$ scans, which served as the reference MBF values in this paper.}
\label{fig_bland_altman}
\end{figure*}

The average regional $K_1$, $V_b$, MBF and MFR values for all 9 normal volunteers are presented in Table \ref{table2}. For all the 9 subjects, $G_S$ produced lower-noise images with lower $K_1$ than the original dynamic frames (an average $17.20\%$ decrease compared to dynamic frames). The proposed $G_{prc}$ improved image contrast with $K_1$ values higher than denoised images. Compared to dynamic frames, $G_{prc}$ lowered the $K_1$ values by $11.20\%$ on average. $G_{prc}$ consistently produced images with lower $V_b$ values, indicating a better separation between the left and right ventricular blood pools (with an average $14.74\%$ decrease compared to dynamic frames). MBF values quantified using the $^{15}\text{O-water}$ were used as the reference in this paper. As presented in Table \ref{table2}, the linear fitting plots in Fig. \ref{fig6}, and the Bland-Altman plots in Fig. \ref{fig_bland_altman}, the proposed method produced images with $^{82}\text{Rb}$ MBFs more consistent with $^{15}\text{O-water}$ MBFs. After applying the proposed simoutaneous denoising and positron range correction method, compared with $^{15}\text{O-water}$ MBFs, the mean MBF differences decrease from 0.431 to 0.088.

\begin{table}[h]
\centering
\caption{Mean $K_1$, $V_b$, MBF, and MFR values for all the 9 normal volunteers acquired on a Siemens mCT PET/CT system at the Yale PET Center. MBF values obtained from the $^{15}\text{O-water}$ scans were used as the reference in this paper. MBF measurements from images reconstructed by the proposed method are better aligned with the MBF values from $^{15}\text{O-water}$ scans. Absolute percentage difference between arterial input functions (AIF) and the image-derived input functions (IDIF) are also included. Proposed method produced images with TACs better matched with AIFs with a overall lower percentage difference.}
\resizebox{\linewidth}{!}{
\begin{tabular}{c|c|c|c|c}
\hline
\multicolumn{5}{c}{\textbf{Siemens mCT PET/CT system}}\\
\hline
    & & $K_1$ &  $V_b$ & MBF       \\ \hline
\multirow{4}{*}{\makecell{ Rest \\ Scans }}
& $^{82}\text{Rb}$ Recon & $0.65\pm0.05$& $0.35\pm0.06$& $1.31\pm0.15$\\ 
\cline{2-5}
 & Denoised & $0.52\pm0.07$ & $0.36\pm0.06$ & $0.90\pm0.22$\\
 \cline{2-5}
  & Denoised+PRC & $0.55\pm0.05$ & $0.31\pm0.06$& $0.98\pm0.15$\\
     \cline{2-5}
  & $^{15}\text{O-water}$ & $1.02\pm0.11$ & $0.28\pm0.08$ & $1.05\pm0.17$\\
  \hline
  \hline
  \multirow{4}{*}{\makecell{ Stress \\ Scans }} & $^{82}\text{Rb}$ Recon & $1.43\pm0.12$& $0.30\pm0.08$& $4.16\pm 0.31$\\ 
\cline{2-5}
 & Denoised & $1.22\pm0.13$& $0.32\pm0.09$& $3.34\pm0.47$\\
 \cline{2-5}
  & Denoised+PRC & $1.33\pm0.11$& $0.25\pm0.08$& $3.79\pm0.35$\\
   \cline{2-5}
  & $^{15}\text{O-water}$ & $3.61\pm0.54$& $0.33\pm0.20$& $3.55\pm0.36$\\

  \hline \hline

  \multirow{4}{*}{\makecell{ MFR }} & $^{82}\text{Rb}$ Recon & \multicolumn{3}{|c}{$3.21\pm0.33$} \\
  \cline{2-5}
  & Denoised & \multicolumn{3}{|c}{$3.76\pm0.38$} \\
  \cline{2-5}
    & Denoised+PRC & \multicolumn{3}{|c}{$3.91\pm0.56$} \\
  \cline{2-5}
      & $^{15}\text{O-water}$ & \multicolumn{3}{|c}{$3.46\pm0.62$} \\
  \hline\hline
  
      & & AUC &  Peak & Tail       \\ \hline
    \multirow{3}{*}{\makecell{ AIF v.s. IDIF\\ (absolute \% difference) }} & $^{82}\text{Rb}$ Recon & $11.09\pm10.46$ & $10.94\pm12.31$ & $9.62\pm10.52$ \\
  \cline{2-5}
  & Denoised & $7.63\pm8.78$ & $9.41\pm12.82$ & $15.18\pm15.53$ \\
  \cline{2-5}
    & Denoised+PRC & $7.58\pm7.93$ & $9.39\pm12.65$ & $9.48\pm10.51$ \\
  \hline
\end{tabular}
}
\label{table2}
\end{table}

\subsection{Generalizability Test}

\begin{figure}[!t]
\centerline{\includegraphics[width=\linewidth]{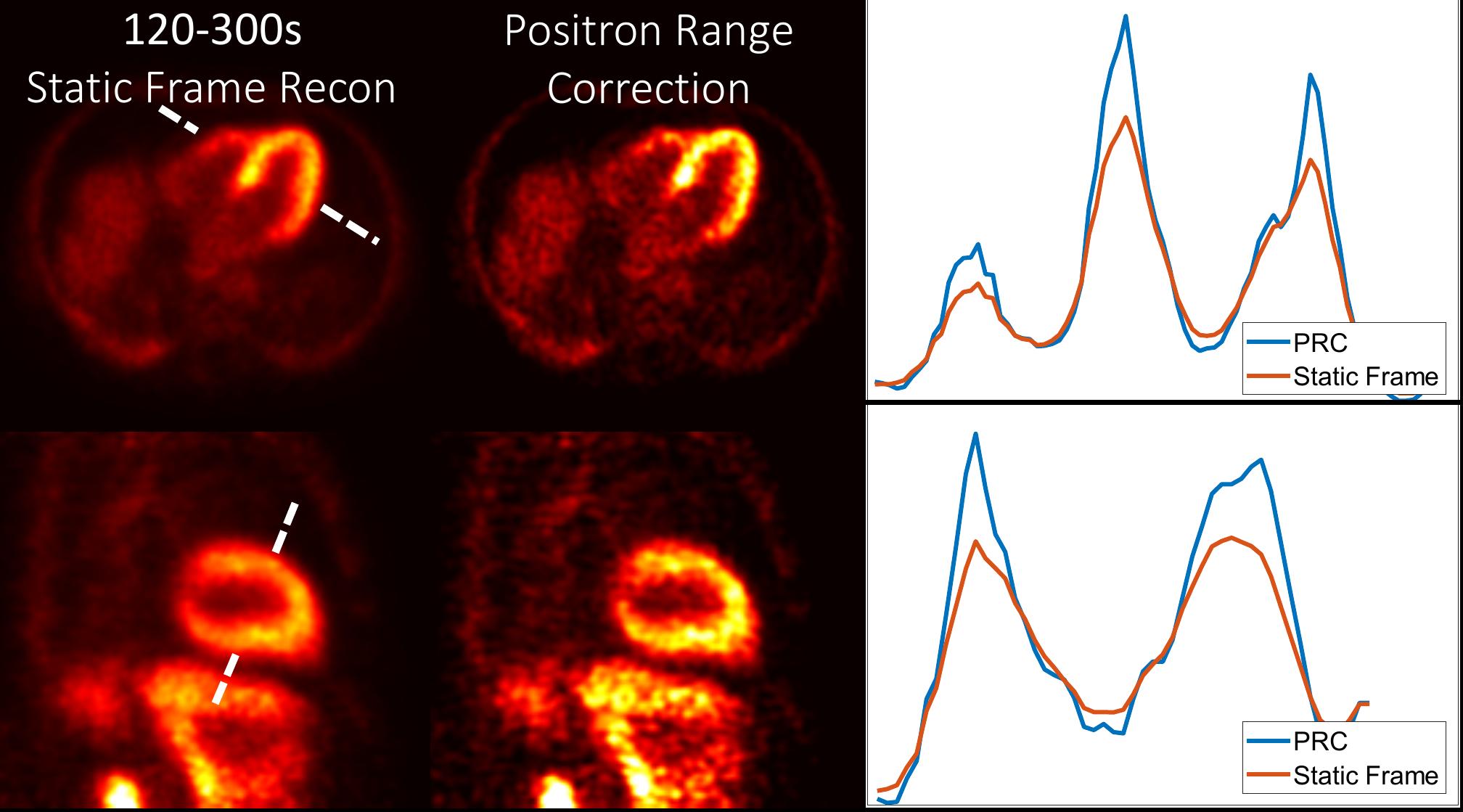}}
\caption{Patient study from the Siemens Vision PET/CT system. The model trained on Siemens mCT data was used to demonstrate the generalizability of the proposed method. Profile plots were generated along the dashed white lines.}
\label{fig8}
\end{figure}

The positron range effect should be independent of the scanner. To evaluate the generalizability of the proposed positron range correction method, we directly apply the trained model on 37 patient scans obtained on a different scanner (Siemens Biograph Vision PET/CT) at University of Manchester Hospital. One patient study is presented in Fig. \ref{fig8}. The proposed positron range correction method produced images with better resolution without further fine-tuning, as validated by the profile plots in Fig. \ref{fig8}.

\begin{figure*}[!t]
\centerline{\includegraphics[width=\linewidth]{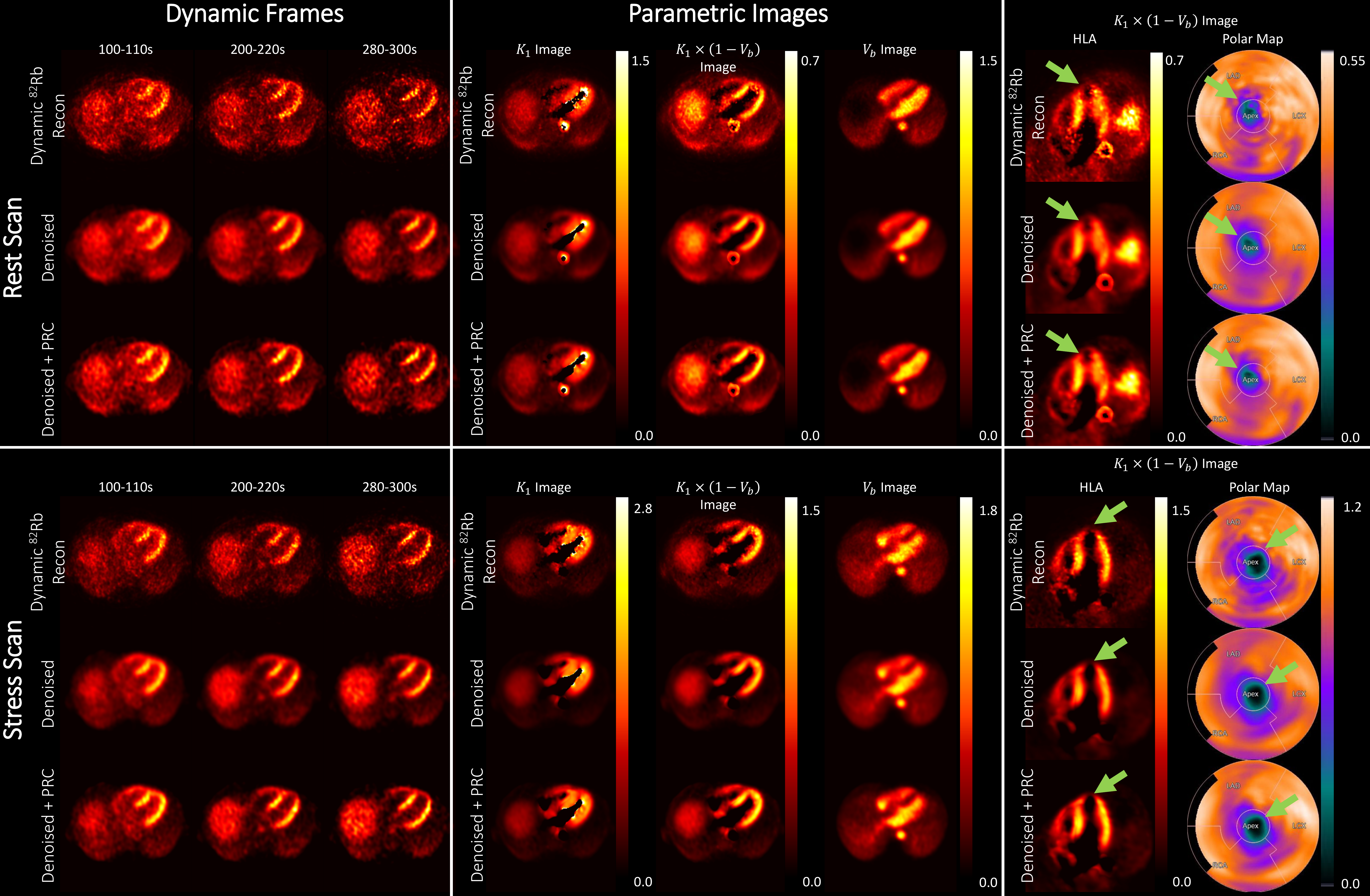}}
\caption{Dynamic frames from a sample patient study acquired on a Siemens Vision PET/CT system at the University of Manchester Hospital. The proposed method was able to generalize to data acquired on a different scanner without additional fine-tuning. Dynamic frames from both rest and stress studies are presented. Note the noise-level differences between early and later frames. The proposed method can generalize to images with different noise levels. $K_1$ images are multiplied by $(1-V_b)$ to remove the artifacts at the boundary of myocardium and blood pool for better visualization (denoted as $K_1\times(1-V_b)$ image). Green arrows in the horizontal long-axis (HLA) images and the polar maps point to the apical perfusion defect in this patient.}
\label{fig9}
\end{figure*}

We also applied both the denoising and positron range correction methods to dynamic data obtained on the Siemens Vision PET/CT system. One sample patient study with an apical defect is presented in Fig. \ref{fig9}, the proposed method produced images with lower noise and higher contrast without additional fine-tuning. Also, dynamic data obtained from the Siemens Vision PET/CT system generally have lower noise due to higher scanner sensitivities. Results in Fig. \ref{fig9} demonstrate the generalizibility of the proposed dynamic denoising method to different noise-levels, tracer distributions, patient populations, and scanners. The superior generalizability of the network could be helpful in clinical translation.

As presented in Fig. \ref{fig9}, the proposed method also produced lower-noise parametric images on Siemens Vision PET/CT system. The corresponding polar maps are also less noisy, making the true apical defect better visualized after applying the proposed method. For the rest scans, the regional apical MBF values are 0.618, 0.364, and 0.474 ml/min/g for the original dynamic frames, output from $G_S$, and output from $G_{prc}$, respectively. These numbers are 1.3961, 0.9635, and 1.1663 ml/min/g for the stress scans. Lower regional MBF values suggest a better defect contrast in this patient study. However, further investigations are needed to demonstrate the clinical potential.

The average regional $K_1$, $V_b$, MBF, and MFR values for all the 37 patient studies are presented in Table \ref{table3}. Similarly, the denoising network $G_S$ produced lower-noise images with lower $K_1$ than the original dynamic frames (a 3.58\% decrease compared with dynamic frames). After applying the positron range correction network $G_{prc}$, the $K_1$ values are close to the original dynamic frames (with only a 0.32\% decrease). $G_{prc}$ consistently produced images with lower $V_b$ values, indicating a better separation between the left and right ventricular blood pools (with an average 12.69\% decrease compared with original dynamic frames).

For patient studies acquired on a Siemens Vision PET/CT system, even though the proposed framework for simultaneous dynamic image denoising and positron range correction ($G_S+G_{prc}$) produced lower-noise images, it does not significantly affect the MBF quantification results compared with MBF values obtained using the original dynamic frames ($p=0.54$). However, the denoising network $G_S$ alone did lower the MBF measurements with statistical significance ($p<0.001$). We suspect that it was because $G_S$ not only reduced image noise but also blurred the images, resulting in overall lower $K_1$ and MBF values.

Using the static frame results, the LV volumes were quantified using the Carimas software \cite{rainio_carimas_2023}. Quantification of LV volume provides prognostic value and serves as a predictive measure of heart health \cite{bravo_reference_2010}. After positron range correction, we observed an increase in LV volume. This is consistent with our expectation as the proposed positron range correction method helps mitigate the positron range blurring, resulting in sharper and more precise LV boundaries. The measured LV volumes are $30.35\pm10.80 \text{ ml}$ and $39.17\pm13.59\text{ ml}$ ($p<0.001$) for static frame inputs and the positron range correction results, respectively.

\begin{table}[h]
\centering
\caption{Mean $K_1$, $V_b$, MBF, and MFR values for all the 37 patient studies acquired on a Siemens Vision PET/CT system at the University of Manchester Hospital.}
\resizebox{\linewidth}{!}{
\begin{tabular}{c|c|c|c|c}
\hline
\multicolumn{5}{c}{\textbf{Siemens Vision PET/CT system}}\\
\hline
    & & $K_1$ &  $V_b$ & MBF       \\
    \hline

\multirow{3}{*}{\makecell{ Rest \\ Scans }}
& $^{82}\text{Rb}$ Recon & $0.690\pm0.146$& $0.296\pm0.061$& $1.450\pm0.499$\\ 
\cline{2-5}
 & Denoised & $0.666\pm0.147$ & $0.290\pm0.058$ & $1.372\pm0.496$\\ 
 \cline{2-5}
  & Denoised+PRC & $0.687\pm0.147$ & $0.256\pm0.059$& $1.441\pm0.505$\\
     \cline{2-5}
     
    \hline\hline
    \multirow{3}{*}{\makecell{ Stress \\ Scans }} & $^{82}\text{Rb}$ Recon & $1.296\pm0.286$& $0.348\pm0.104$& $3.637\pm 1.069$\\ 
\cline{2-5}
 & Denoised & $1.254\pm0.304$& $0.352\pm0.105$& $3.471\pm1.119$\\
 \cline{2-5}
  & Denoised+PRC & $1.299\pm0.317$& $0.314\pm0.109$& $3.627\pm1.136$\\
   \cline{2-5}
  \hline \hline

  \multirow{3}{*}{\makecell{ MFR }} & $^{82}\text{Rb}$ Recon & \multicolumn{3}{|c}{$2.614\pm0.653$} \\
  \cline{2-5}
  & Denoised & \multicolumn{3}{|c}{$2.640\pm0.691$} \\
  \cline{2-5}
    & Denoised+PRC & \multicolumn{3}{|c}{$2.630\pm0.706$} \\
  \hline

\end{tabular}
}
\label{table3}
\end{table}

\subsection{Comparison with Other Denoising Methods}
\label{sec3.4}
Deep learning for medical image denoising has been widely investigated in the literature \cite{xu_200x_2017, zhou2020supervised, ouyang_ultra-low-dose_2019, zhou2023fast, zhou_federated_2023, zhou2023fedftn, gong_pet_2019, onishi_anatomical-guided_2021, xie_unified_2024, liu_pet_2022}. Even though existing methods cannot be directly applied to dynamic $^{82}\text{Rb}$ PET denoising due to the limitations mentioned previously,  we believe comparisons with other related methods will still be beneficial to show the effectiveness of the proposed method.

In this subsection, the proposed denoising neural network (i.e., $G_S$) is compared with the following methods:

\begin{enumerate}
  \item The Unified Noise-aware Network (UNN) \cite{xie_unified_2024}. UNN was chosen because: (1) similar to the proposed $G_S$, UNN also achieves noise-aware denoising; (2) and it was among the top 10 winning methods in the Ultra Low-dose PET Imaging Challenge held at the 2022 IEEE Medical Imaging Conference (IEEE MIC) and the 2022 International Conference on Medical Image Computing and Computer Assisted Intervention (MICCAI) \footnote{https://ultra-low-dose-pet.grand-challenge.org/leaderboard/}.
  \item Diffusion model for PET image denoising introduced in this paper \cite{gong_pet_2024}. This method was chosen due to the recent popularity of diffusion model. Recently, diffusion models have become the new state-of-the-art generative models \cite{Croitoru.etal2023}. They are capable of generating high-quality samples from Gaussian noise input, and have demonstrated strong potential for low-dose PET imaging. To denoise the entire 4D dynamic series in a reasonable amount of time, the Denoising Diffusion Implicit Models (DDIM) \cite{song_denoising_2022} sampling was implemented for comparison (denoted as DDIM-PET in this paper).
  \item The Noise2Void method for PET image denoising introduced in this paper \cite{song_noise2void_2021}. This method was chosen as it also achieves PET image denoising without paired inputs/labels and it is directly related to the proposed method in this paper.
  \item To show the effectiveness of the proposed dynamic convolutional strategy (illustrated in Fig.\ref{fig2}), the proposed method without this component was included as an ablation study (denoted as $G_{S\text{ No Dyn Conv}}$ in this paper).
\end{enumerate}
Note that since both the UNN and DDIM-PET requires paired inputs/labels for network training, which is not available for dynamic $^{82}\text{Rb}$ denoising, they were trained using 90 patient studies with \textsuperscript{18}F-FDG tracer acquired at the Yale-New Haven Hospital. Another 10 subjects were included for validation purpose. These patient studies were acquired using a Siemens Biograph mCT PET/CT system. To simulate the varying noise levels in 4D $^{82}\text{Rb}$ dynamic series, images with 5\%, 10\%, and 20\% low-count levels were reconstructed through listmode rebinning. The trained model was directly applied for $^{82}\text{Rb}$ dynamic denoising.

The Noise2Void \cite{song_noise2void_2021} method and $G_{S\text{ No Dyn Conv}}$ do not require paired inputs/labels. They were trained and tested in the same way as described previously in this paper.

\begin{figure*}[!t]
\centerline{\includegraphics[width=\linewidth]{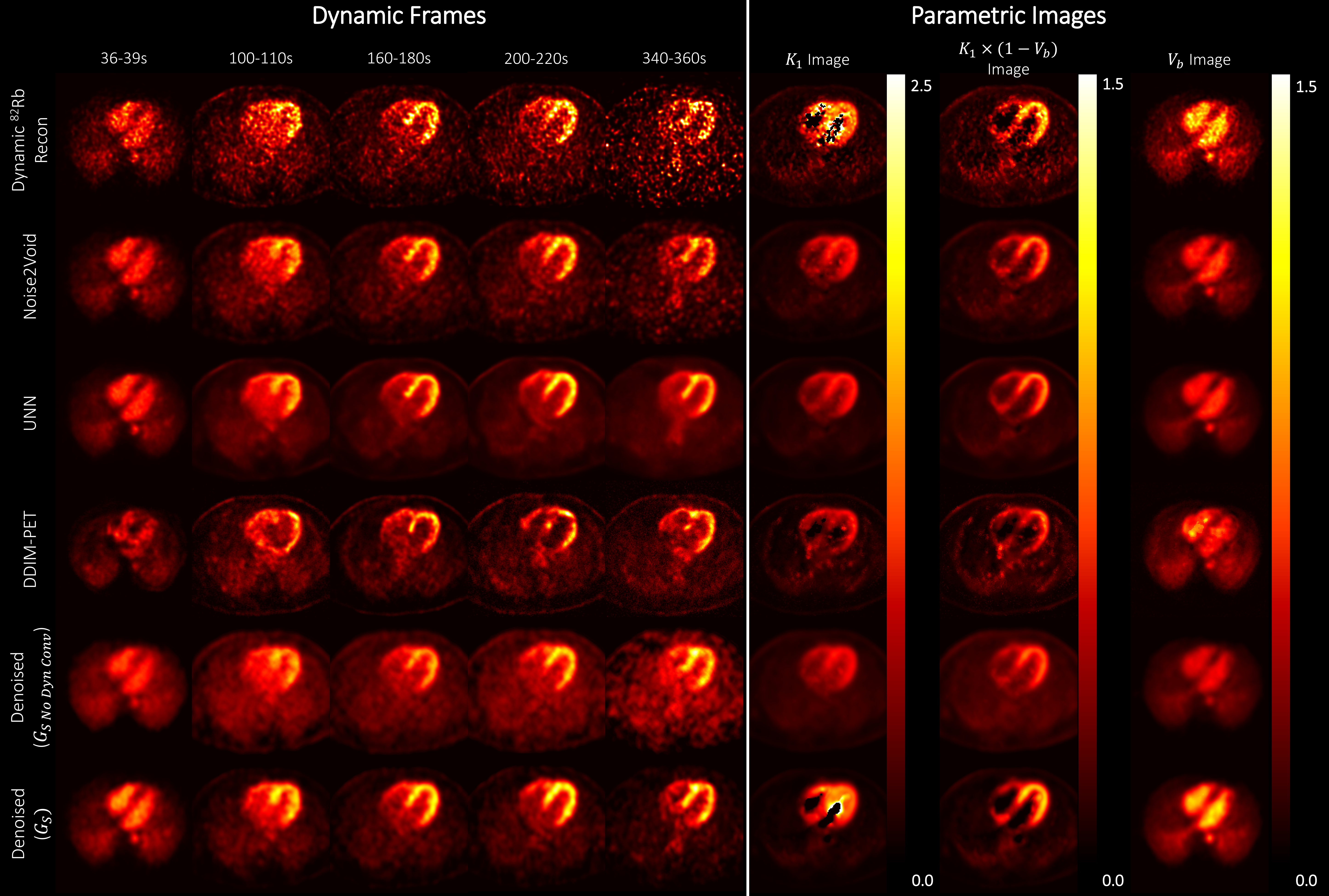}}
\caption{Denoised dynamic frames and the corresponding parametric images generated using different methods from a sample normal volunteer study. $K_1$ images are multiplied by $(1-V_b)$ to remove the artifacts at the boundary of myocardium and blood pool for better visualization (denoted as $K_1\times(1-V_b)$ image).}
\label{fig10}
\end{figure*}

Sample denoised images using different methods are presented in Fig. \ref{fig10}. Since the Noise2Void method was trained using images with varying image noise levels and tracer distributions, without any noise-aware or temporal-aware strategy, it is not able to produce optimal denoised results across different dynamic frames. Compared to images generated using the proposed denoising method $G_S$, Noise2Void produced images with less uniform myocardium for this normal volunteer study. For the study shown in Fig. \ref{fig10}, the standard deviations of voxel values in the myocardium VOI for all the dynamic frames are $1.82\times10^4\text{Bq/ml}$, $1.43\times10^4\text{Bq/ml}$, and $1.16\times10^4\text{Bq/ml}$ for the original dynamic frames, outputs from Noise2Void, and outputs from the proposed denoising method $G_S$, respectively. Lower standard deviation represents a more uniform myocardium, which is desirable for a normal volunteer study. Similarly, without the proposed dynamic convolutional strategy to achieve noise- and temporal-awareness, the network $G_{S\text{ No Dyn Conv}}$ produced images with higher blood-pool and lung activities. We suspect that early frame images with higher background activities affect late frames denoised results in the $G_{S\text{ No Dyn Conv}}$ network, leading to a overall higher $V_b$ values (with an average 23.31\% increase compared to original dynamic frames).

Even though UNN achieved noise-aware denoising and produced visually-promising denoised results across dynamic frames, it introduced undesired smoothness to the images, leading to overall lower $K_1$ values (with an average 11.11\% decrease compared to original dynamic frames). Since the UNN network was trained using $^{18}\text{F-FDG}$ studies, it did not generalize well to images acquired with a different tracer.

DDIM-PET produced images with distorted myocardium, leading to higher variances of $K_1$ and $V_b$ values as presented in Table \ref{table4}, especially for the stress scans. We suspect it was because the stochastic nature of diffusion model and the generalizability issue as it was also trained using $^{18}\text{F-FDG}$ studies.

To show the improvement in MBF quantification, Table \ref{table4} presents the mean absolute differences between the MBF measurements obtained from different denoised images and the corresponding $^{15}\text{O-water}$ scans. The proposed denoising method $G_S$ produced images with MBF measurements closest to that quantified using $^{15}\text{O-water}$ scans.

\begin{table}[h]
\centering
\caption{Comparison between different denoising methods. Mean $K_1$, $V_b$, MBF, and MFR values for different methods for all the 9 normal volunteers acquired on a Siemens mCT PET/CT system at the Yale PET Center. $G_{S}$ represents the proposed denoising network. Using MBF values obtained from $^{15}\text{O-water}$ scans as reference, the mean absolute differences (MAE) between MBF measurements from different denoised images and the corresponding $^{15}\text{O-water}$ scans are included in this table. The MBF measurements with the lowest mean differences are marked in bold.}
\resizebox{\linewidth}{!}{
\begin{tabular}{c|c|c|c|c|c}
\hline
\multicolumn{6}{c}{\textbf{Siemens mCT PET/CT system}}\\
\hline
    & & $K_1$ &  $V_b$ & MBF  & MAE     \\
    \hline

\multirow{6}{*}{\makecell{ Rest \\ Scans }}
& $^{82}\text{Rb}$ Recon & $0.65\pm0.05$& $0.35\pm0.06$& $1.31\pm0.15$ & $0.26\pm0.19$\\ 
\cline{2-6}
& UNN & $0.56\pm0.12$ & $0.39\pm0.07$ & $1.02\pm0.38$& $0.31\pm0.37$\\
\cline{2-6}
& DDIM-PET & $0.85\pm0.12$ & $0.42\pm0.09$ & $2.10\pm0.42$& $1.05\pm0.39$\\
\cline{2-6}
& Noise2Void & $0.50\pm0.08$ & $0.38\pm0.07$ & $0.85\pm0.23$& $0.26\pm0.22$\\
\cline{2-6}
& $G_{S\text{ No Dyn Conv}}$ & $0.45\pm0.06$ & $0.45\pm0.06$ & $0.70\pm0.16$& $0.35\pm0.21$\\
\cline{2-6}
 & Denoised ($G_S$) & $0.52\pm0.07$ & $0.36\pm0.06$ & $0.90\pm0.22$ & $\boldsymbol{0.22\pm0.12}$\\
 \cline{2-6}

     \hline\hline
\multirow{6}{*}{\makecell{ Stress \\ Scans }} & $^{82}\text{Rb}$ Recon & $1.43\pm0.12$& $0.30\pm0.08$& $4.16\pm 0.31$ & $0.61\pm0.40$\\ 
\cline{2-6}
& UNN & $1.30\pm0.20$ & $0.28\pm0.07$ & $3.66\pm0.75$ & $0.71\pm0.41$\\
\cline{2-6}
& DDIM-PET & $1.94\pm1.40$ & $0.48\pm0.22$ & $4.54\pm1.29$& $1.40\pm0.75$\\
\cline{2-6}
& Noise2Void & $1.13\pm0.16$ & $0.39\pm0.09$ & $3.01\pm0.59$& $0.73\pm0.47$\\
\cline{2-6}
& $G_{S\text{ No Dyn Conv}}$ & $1.01\pm0.14$ & $0.45\pm0.06$ & $2.58\pm0.50$& $0.98\pm0.37$\\
\cline{2-6}
 & Denoised ($G_S$) & $1.22\pm0.13$& $0.32\pm0.09$& $3.34\pm0.47$ & $\boldsymbol{0.38\pm0.15}$\\
 \cline{2-6}
  \hline \hline

  \multirow{6}{*}{\makecell{ MFR }} & $^{82}\text{Rb}$ Recon & \multicolumn{4}{|c}{$3.21\pm0.33$} \\
  \cline{2-6}
  & UNN & \multicolumn{4}{|c}{$3.83\pm1.00$} \\
  \cline{2-6}
  & DDIM-PET & \multicolumn{4}{|c}{$2.15\pm0.49$} \\
  \cline{2-6}
  & Noise2Void & \multicolumn{4}{|c}{$3.80\pm1.27$} \\
  \cline{2-6}
  & $G_{S\text{ No Dyn Conv}}$ & \multicolumn{4}{|c}{$3.76\pm0.84$} \\
  \cline{2-6}
  & Denoised ($G_S$) & \multicolumn{4}{|c}{$3.80\pm0.38$}\\
  \hline

\end{tabular}
}
\label{table4}
\end{table}

%% file: sec/conclusion.tex
\section{Discussion and Conclusion}
\label{Conclusion}

Cardiovascular disease remains as the leading cause of death worldwide \cite{tsao_heart_2023}, and tracer kinetic modeling with $^{82}\text{Rb}$ cardiac PET have shown prognostic values for the assessment of cardiovascular diseases \cite{murthy_clinical_2018} (especially the quantification of MBF and MFR). In this work, we present a deep learning approach to address two of the physical factors that negatively affect $^{82}\text{Rb}$ cardiac PET image quality and quantitative accuracy. First, the short half-life results in noisy reconstructions of dynamic frames and parametric images, and supervised labels are not available due to tracer decay. Noise levels also vary among different dynamic frames. Here, we proposed a self-supervised method to achieve noise-aware image denoising to account for these issues. The proposed method produced consistent denoised results regardless of the input noise levels, tracer distributions, and even different scanners in different medical institutions. Second, the longer positron range of $^{82}\text{Rb}$ limits the image spatial resolution. Here, we proposed a self-supervised method to approximate the inverse of the Monte-Carlo-simulated positron range distributions to achieve positron range correction. The proposed method produced images with higher contrast and better recovery of subtle cardiac features (e.g. papillary muscles).
The proposed method also produced lower noise parametric images, which may facilitate the utilization of parametric imaging in clinical settings \cite{wang_pet_2020}. As presented in the results section, the proposed method also produced $V_b$ images with better separation between left and right ventricular blood pools. This may allow better quantification of the MBF of septal wall and the intramyocardial blood volume \cite{xie_segmentation-free_2023, mohy-ud-din_quantification_2018} for the diagnosis of coronary micro-vascular diseases, a major subset of ischemic heart disease.

To the best of our knowledge, this work is the first attempt to use a deep-learning approach to achieve both noise reduction and positron range correction for $^{82}\text{Rb}$ cardiac PET imaging. 

In this preliminary study, we demonstrated the feasibility of using a deep learning approach to achieve simultaneous dynamic image denoising and positron range correction for $^{82}\text{Rb}$ cardiac PET imaging using a self-supervised method. The proposed method potentially improved the quantification of myocardium blood flow as validated against $^{15}\text{O-water}$ scans as well as radioactivities quantified from arterial blood samplings on normal volunteer studies. Since we do not have access to the diagnostic comments for the patient studies, the main limitation of this work is the lack of clinical validation. In the future, we plan to evaluate the proposed method using patient data with invasive hemodynamics to further investigate the clinical potential of this work. We believe the proposed method for self-supervised noise-aware dynamic image denoising could be easily extended to other medical imaging applications in which paired labels are not easily obtained. Also, the proposed method only considers a uniform kernel for positron range correction. A method to consider heterogeneous kernels is required for general-purpose positron range correction for different organs or total-body PET scans.

\section*{Acknowledgments}
This work was supported by NIH under Grants R01EB025468, R01HL154345, R01HL169868, R01CA275188, and a research contract from Siemens Healthineers.